\RequirePackage{fix-cm}
\documentclass[twocolumn]{svjour3}          
\smartqed  
\usepackage{graphicx}
\usepackage{amssymb}
\usepackage{graphicx}
\usepackage{authblk}
\usepackage{natbib}
\usepackage[english]{babel}
\addto{\captionsenglish}{%
    \renewcommand{\bibname}{References}
}
\usepackage{indentfirst}
\usepackage{rotating}
\usepackage{lscape}
\usepackage{float}
\usepackage{epstopdf, epsfig}
\usepackage[labelfont=bf,textfont=normalfont,singlelinecheck=off,justification=raggedright]{}
\usepackage{array}
\usepackage{makecell}

\def\makeheadbox{{%
\hbox to0pt{\vbox{\baselineskip=10dd\hrule\hbox
to\hsize{\vrule\kern3pt\vbox{\kern3pt
\hbox{ }
\kern3pt}\hfil\kern3pt\vrule}\hrule}%
\hss}}}
\begin{document}

\title{Internal flows of ventilated partial cavitation
}


\author{Kyungduck Yoon \and   
        Shijie Qin \and
        Siyao Shao \and
        Jiarong Hong
}

\authorrunning{K. Yoon, S. Qin, S. Shao, and J. Hong} 

\institute{Kyungduck Yoon \and
        Siyao Shao \and
        Jiarong Hong \at
        St. Anthony Falls Laboratory, University of Minnesota, Minneapolis, MN 55414, USA
        \and
        Kyungduck Yoon \and
        Siyao Shao \and
        Jiarong Hong \at
        Department of Mechanical Engineering, University of Minnesota, Minneapolis, MN 55414, USA
        \and
        Shijie Qin \at
        Ocean College, Zhejiang University, Zhoushan 316021, China
}

\date{Received: date / Accepted: date}

\maketitle

\begin{abstract}
Our study provides the first experimental investigation of the internal flows of ventilated partial cavitation (VPC) formed by air injection behind a backward-facing step. 
The experiments are conducted using flow visualization and planar particle image velocimetry (PIV) with fog particles for two different cavity regimes of VPC, i.e., open cavity (OC) and two-branch cavity (TBC), under various range of free stream velocity ($U$) and ventilation rates ($Q$). 
Our experiments reveal similar flow patterns for both OC and TBC, including forward flow region near the air-water interface, reverse flow region, near-cavitator vortex, and internal flow circulation vortex. 
However, OC internal flow exhibits highly unsteady internal flow features, while TBC internal flow shows laminar-like flow patterns with a Kelvin-Helmholtz instability developed at the interface between forward and reverse flow regions within the cavity. 
Internal flow patterns and the unsteadiness of OC resemble those of turbulent flow separation past a backward-facing step ($BFS$ flow), suggesting a strong coupling of internal flow and turbulent external recirculation region for OC. Likewise, internal flow patterns of TBC resemble those of laminar $BFS$ flow, with the presence of unsteadiness due to the strong velocity gradient across the forward-reverse flow interface. 
The variation of the internal flow upon changing $U$ or $Q$ are further employed to explain the cavity regime transition and the corresponding change of cavity geometry. 
Our study suggests that the ventilation control can potentially stabilize the cavity in the TBC regime by delaying its internal flow regime transition from laminar-like to highly unsteady. 
\end{abstract}

\section{Introduction}
\label{intro}

Ventilated partial cavitation (VPC) is an air pocket that is artificially formed by gas injection behind a flow separation device or cavitator, in a liquid flow. Such phenomenon has been applied to air lubrication on ship operation for friction drag reduction \citep{ceccio2010ARFM}, also known as partial cavity drag reduction (PCDR). When properly designed, PCDR has been shown to achieve up to 95\% drag reduction for the covered area and 20\% energy savings \citep{makiharju2012IJNAOE}.

To utilize this promising technique on energy savings for ship operations, extensive work has been done on PCDR, mainly focusing on the cavitator design and establishing a relationship between relevant flow conditions and ventilation cost. For instance, \citet{amromin2006JFE} and \citet{kopriva2008JFE} measured the lift and drag coefficients on a hydrofoil cavitator to validate the effectiveness of cavitation theory for drag reduction schemes. \citet{makiharju2010SNH28th} also used imaging techniques on cavity closure to determine the cavity shape, growth rate and collapse rate, closure oscillation, and closure type and observed the effect of each flow condition on the ventilation cost. To ensure the industrial applicability of PCDR, \citet{makiharju2013JFM} later conducted both large- and small-scale experiments to examine the scaling factors of relevant flow parameters to the required ventilation rate for a partial cavity to be sustained under various flow parameters.

In addition to studies on PCDR, studies on cavity shape and topology have also been conducted. 
\citet{laberteaux2001JFM} used particle image velocimetry (PIV) and particle streak photography to investigate the effects of external flow on cavity geometry and air entrainment of a natural partial cavity formed behind a wedge. \citet{barbaca2017IJMF} and \citet{barbaca2018IJMF} used bottom view imaging of a VPC that formed behind a 3D and 2D fence type cavitator, respectively, to classify the cavity topology based on gas leakage mechanisms. They also performed pressure measurements to calculate the cavitation number in each regime. 
Most recently, using high-speed imaging on the side and bottom views, \citet{qin2019IJMF} investigated VPC under various flow conditions and ventilation rates and classified cavity regimes based on gas leakage mechanisms and explained the transitions between regimes. 
They further suggested the importance of understanding the role of internal flow in a gas leakage mechanism. 
They pointed out that gas leakage occurs as the ventilated gas inside the cavity is dragged by the internal boundary layer and dispersed into bubbles by the external recirculating vortex that is formed behind the closure of the cavity. 
However, no prior studies have provided the internal flow characterization of VPC. 
Thus, the purpose of this study is to experimentally investigate the internal flow of VPC to further understand the ventilation physics of VPC.

In contrast to the dearth of internal flow studies in VPC, only two studies have experimentally examined the internal flow of ventilated supercavitation. 
\citet{wang2018IJMF} conducted internal flow PIV on a 2D cylinder cavity by seeding fluorescent-painted glass micro-spheres and acquired time-averaged velocity field. 
However, the spatial resolution of the acquired flow field was not sufficient enough to reveal flow structures that occur in the instantaneous internal flow field. 
\citet{wu2019JFM} used the ray tracing mapping method that enabled distortion correction of a curved surface to perform high-resolution internal flow PIV with fog particles seeded. 
The resolution of their experiments were sufficient enough to observe the coherent structures and to provide quantitative measurements for supercavity with different closure types. 
They not only provided information on the internal boundary layer that was not available in the literature but further provided a detailed explanation on cavity evolution, closure formation, gas leakage mechanism based on the coupling of internal and external flows.

Employing a similar approach to that of \citet{wu2019JFM}, the current study presents an experimental investigation on the internal flow of VPC using flow visualization and PIV. This study aims to characterize the internal flow field under distinct cavity regimes and investigate the connection of the internal flow with general cavity behaviors (i.e., cavity length growth and cavity regime). The remaining sections are listed as follows: Section~\ref{sec:Methods} provides detailed description of the experimental methods. The results from our study are presented in Section~\ref{sec:Results}. Specifically, Section~\ref{sec:OC}--\ref{sec:TBC} reports the results of the internal flow visualization and the mean flow field of each cavity regime. A comparison is made between the VPC internal flow and the single phase flow past a backward-facing step in Section~\ref{sec:Analogy} to further explain the cavity behavior upon changes in flow parameters and regime transitions. Section~\ref{sec:Conclusions} provides a summary of the current study.

\section{Experimental Methodology}\label{sec:Methods}
\begin{figure*}[ht]
  \centering
  \includegraphics[width=1\textwidth]{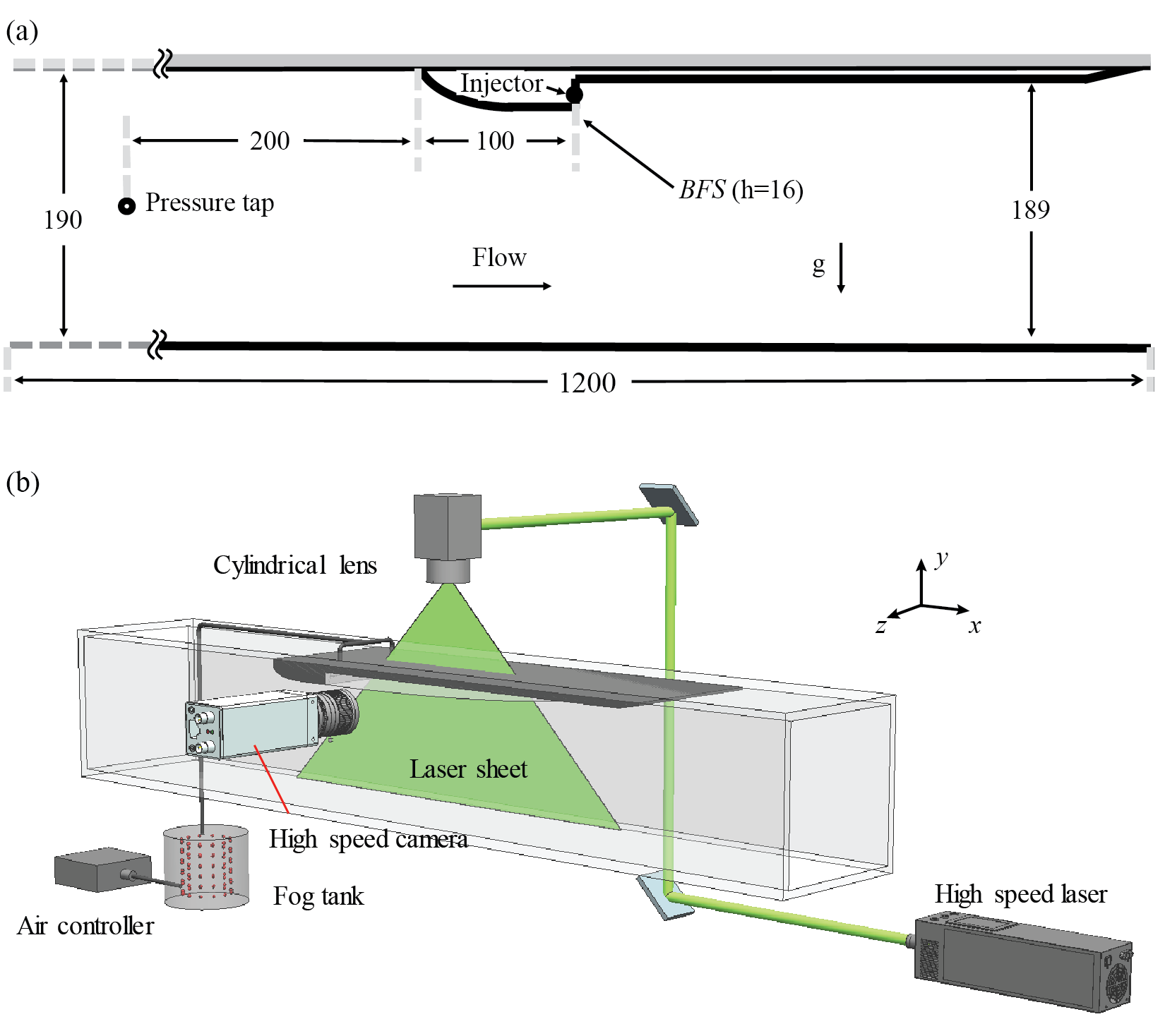}
  \caption{Schematics of (a) the test section with backward-facing step (BFS) cavitator mounted on the ceiling, and (b) the experimental setup. During the experiments, water flows in $x$ direction. All dimensions are in $\mathrm{mm}$.}
\label{fig1}
\end{figure*}
\begin{table*}
    \caption{Experimental conditions of VPC with different cavity regimes. $w$ stands for the width of the test section.}
    \label{table1}
    \begin{tabular}{ l l l l l l }
    \hline\noalign{\smallskip}
      \thead{Case\\No.} & \thead{Free stream velocity $U$\\$\mathrm{m/s}$} & \thead{Ventilation rate $Q$\\$\textrm{SLPM}$} & \thead{Froude number based on h\\$Fr=U/{\sqrt{gh}}$} & \thead{Ventilation Coefficient \\$C_{Q}=Q/Uhw$} & \thead{Cavity regime} \\
      \hline\noalign{\smallskip}
      $1$ &  $1.20$  & $0.7$ & $3.03$ & $3.20\times 10^{-3}$ & OC  \\
      $2$ &  $1.20$  & $0.9$ & $3.03$ & $4.11\times  10^{-3}$ & OC  \\
      $3$ &  $0.65$  & $1.0$ & $1.64$ & $8.44\times  10^{-3}$ & TBC  \\
      $4$ &  $0.65$  & $5.0$ & $1.64$ & $4.22\times  10^{-2}$ & TBC  \\
      $5$ &  $0.80$  & $1.0$ & $2.02$ & $6.85\times  10^{-3}$ & TBC  \\
      $6$ &  $0.80$  & $5.0$ & $2.02$ & $3.43\times  10^{-2}$ & TBC  \\

\hline\noalign{\smallskip}
    \end{tabular}
    
\end{table*}

\begin{table}[t]
\caption{Spatial and temporal resolution of the experiment.}
\label{table2}
    \begin{tabular}{ l l l l l }
    \hline\noalign{\smallskip}
      \thead{Case No.} & \thead{Spatial resolution} & \thead{fps (Hz)} & \thead{Sampling time (s)} \\
      \hline\noalign{\smallskip}
      $1$   & $56.82\  \mathrm{\mu m/pixel}$ & $2500$ & $1.64$ \\
      $2$  & $56.82\  \mathrm{\mu m/pixel}$ & $2500$ & $1.64$ \\
      $3$  & $32.79\  \mathrm{\mu m/pixel}$ & $2000$ & $2.05$ \\
      $4$   & $32.79\  \mathrm{\mu m/pixel}$ & $2000$ & $2.05$ \\
      $5$   & $32.79\  \mathrm{\mu m/pixel}$ & $2500$ & $1.64$ \\
      $6$   & $32.79\  \mathrm{\mu m/pixel}$ & $2500$ & $1.64$ \\
\hline\noalign{\smallskip}
    \end{tabular}
    
\end{table}

\subsection{Experimental facility and setup}
The experiments are conducted in the high-speed cavitation water tunnel at Saint Anthony Falls Laboratory (SAFL), University of Minnesota. The tunnel is a closed recirculating tunnel with a large volume dome-shaped settling chamber designated for fast bubble removal during ventilation experiments. The dimension of the test section is $1.20\ \mathrm{m} \times 0.19 \ \mathrm{m} \times 0.19 \ \mathrm{m}$ (length, height, and width), and the bottom and two side windows of the test section are made up of Plexiglas for optical access. The ceiling of the test section is also transparent for the PIV experiment. A detailed description of the experimental facility can be found in \cite{karn2016JFM}. The 3D printed backward-facing step cavitator with a height of $16 \  \mathrm{mm}$ is installed on the ceiling of test section as shown in Fig.~\ref{fig1}a, which leads to an expansion ratio ($ER$) of $1.09$ and an aspect ratio of $11.9$. VPC is generated by ventilating gas behind a backward-facing step cavitator during the operation of the water tunnel. 

Figure \ref{fig1}b depicts the experimental setup for the PIV experiment. The transparent center portion of the test section ceiling enables the projection of laser sheet from the top in order to ensure that the laser sheet is not distorted by the air-water interface. The laser sheet thickness is calibrated to be $1\pm 0.2 \ \mathrm{mm}$. Particles generated from the Rosco 1700 fog machine act as a tracer that are first filled in 42 gallons for a chamber where the ventilation line is connected so that the particles are laden in the air and injected into the test section. The mass flow rate of the ventilated air is controlled by an Omega Engineering FMA-2609A mass flow controller that controls up to 55 SLPM with the uncertainty within $\pm0.2\% FS$. The fog particle size is estimated to be $29\pm8 \ \mathrm{\mu m}$. A more detailed explanation on the VPC setup can be found in \cite{qin2019IJMF}. The seeding procedures and the calculations of the particle size and laser sheet thickness are detailed in \cite{wu2019JFM}. A Photonics DM30-527 Nd/YLF high-speed laser (maximum pulse energy of $30\  \mathrm{mJ/pulse}$) is synchronized with the time resolution of Photron APX-RS high-speed camera (full sensor size of $1024\times1024$ pixels) to perform time-resolved PIV measurements.

\subsection{Selection of experimental conditions}
Among the four cavity regimes (i.e., foamy cavity, transition cavity, open cavity, and two-branch cavity) reported by \cite{qin2019IJMF}, open cavity (OC) and two-branch cavity (TBC) are selected for the internal flow measurement. 
It is worth noting that the internal flow measurement is only viable for the OC and TBC regimes due to the presence of the large air pocket with a smooth air-water interface whereas the other cavity regimes consist of dispersed bubbles.

Table~\ref{table1} shows the experimental conditions for the VPC internal flow measurement. Experimental conditions are selected using the cavity regime map of \cite{qin2019IJMF} to cover a wide range of flow and ventilation settings for OC and TBC regimes. 
The cavity with various free stream velocities ($U$) and ventilation rates ($Q$) are examined to evaluate the effect of ventilation coefficients ($C_Q$) on the internal flow. 
All experimental conditions are referred to by their case numbers. As shown in Table~\ref{table1}, for OC, we only investigated the experimental conditions with different $Q$ that is precisely controlled (in precision of two decimal places) by the proportional-integral-derivative control algorithm embedded in the mass flow controller. 
In the current experiment, the investigation of OC at higher Froude number ($Fr$) is limited by the substantial cavity length fluctuation and interface distortion occurring with a further increase of $Fr$ which prohibits accurate internal flow PIV measurements.

For TBC, various sets of $C_Q$ are chosen to examine its effect on the internal flow under different $U$ and $Q$ with the remaining parameters fixed.
However, it should be noted that $Fr$ for the TBC regime are selected to be as low as possible since the cavity length of TBC is in the order of $Fr^{2}$ and that the cavity extends beyond the test section for $Fr$ over 3.4 \citep{qin2019IJMF}. 
\begin{figure*}
  \centering
  
  \includegraphics[width=1\textwidth]{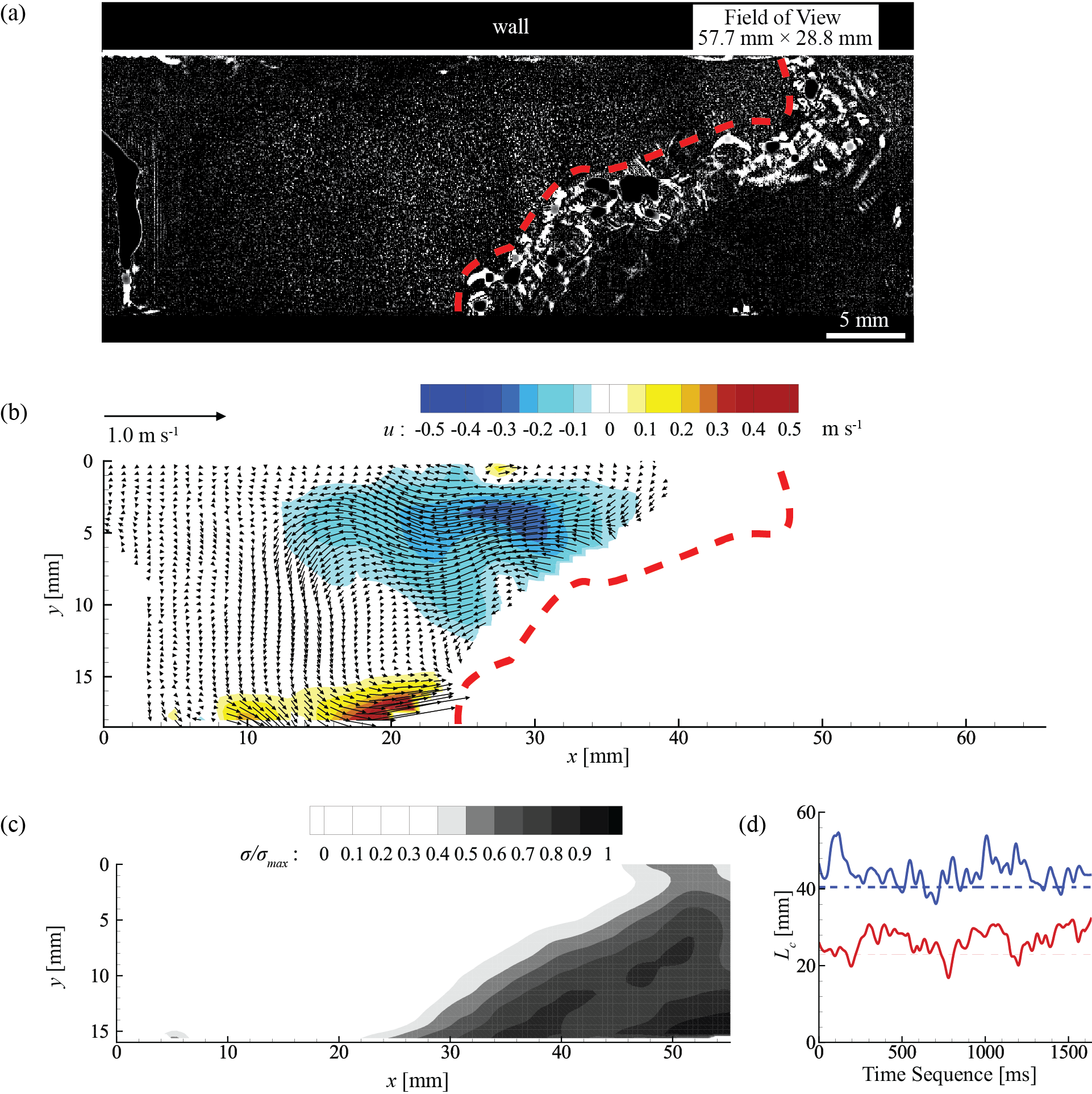}\\
  \caption{Samples of (a) the tracer field image and (b) the corresponding instantaneous velocity vector field superimposed with the streamwise velocity contour map showing the internal flow measurements for VPC in the OC regime ($Fr=3.0$ and $C_Q=3.20\times 10^{-3}$). The red dotted lines of (a) and (b) mark the location of the air-water interface of an instantaneous sample at the closure. For clarity, only one in two vectors is plotted in both directions. The internal flow region of OC is generated by removing the spurious vectors beyond the closure by setting the upper threshold based on (c) the normalized standard deviation of velocity vectors. (d) 85\% of the mean cavity length is chosen as a threshold to ensure that the internal flow field is not affected by unstable closure. The cavity length fluctuations of case 1 and 2 are colored in red and blue lines, respectively. Dotted lines correspond to 85\% of the mean values.}
\label{fig2}
\end{figure*}

\subsection{Measurement settings and data processing}
\begin{figure*}
  \centering
  
  \includegraphics[width=1\textwidth]{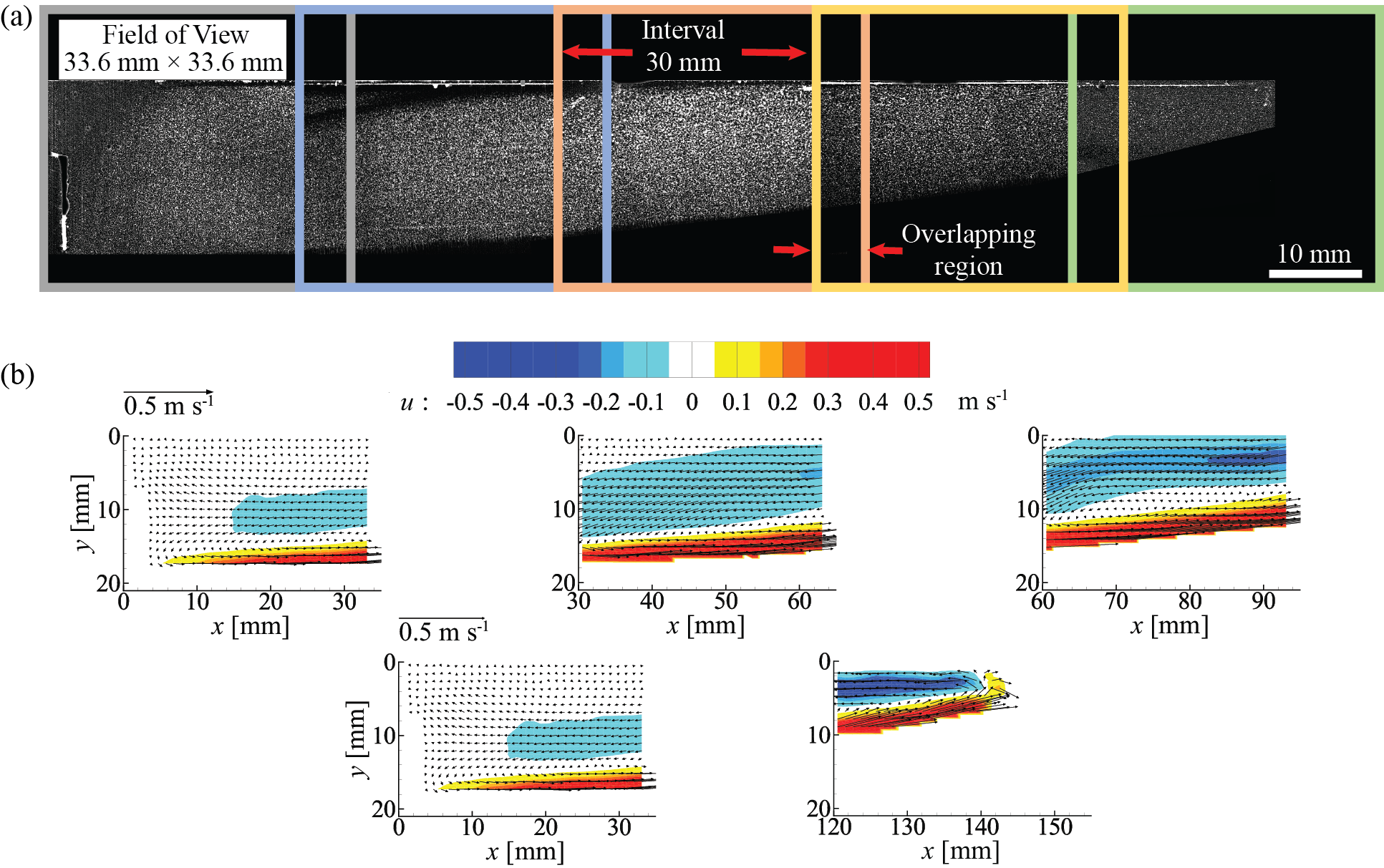}
  
  \caption{Samples of (a) tracer field images and (b) the corresponding instantaneous velocity vector fields superimposed with streamwise velocity contour maps showing the internal flow measurements for VPC in the TBC regime ($Fr=1.6$ and $C_Q=8.44\times 10^{-3}$). For clarity, only one in two vectors is plotted in both directions.}
\label{fig3}
\end{figure*}

We conducted PIV measurements on the streamwise wall-normal plane ($xy$-plane) at the center spanwise ($z$) location of the channel. The PIV recording settings for each case is summarized in Table~\ref{table2}. 
For each PIV case, the corresponding cavity length is measured using the calibrated cavity images. 
It should be noted that the cavity length defined for OC is the distance from the cavitator to the streamwise location where air entrainment occurs, which is uniform along the spanwise ($z$) direction when time-averaged. The cavity length of TBC is defined as the distance from the cavitator to the bifurcation point, since TBC has a bifurcation point at the center $xy$-plane with two branches near the wall of the test section that are elongated downstream. 

Figure~\ref{fig2}a shows a sample image of the tracer field illustrating the measurement field of view (FOV) and image quality. The corresponding instantaneous velocity field is presented in Fig.~\ref{fig2}b, illustrating the internal flow field measured from the PIV calculation. For clarity, only one in two vectors are plotted in both $x$ and $y$ directions. 
The FOV for the OC internal flow PIV (1024$\times$512 pixels) is chosen to capture the entire flow field at once. Tracer images are processed using an adaptive multi-pass cross-correlation algorithm in \emph{LaVision DaVis} 8.2 to obtain instantaneous velocity vector fields. 
The interrogation window starts from 96$\times$96 pixels with a 50\% overlap, decreasing to 32$\times$32 pixels with 75\% overlap. The resulting parameters yield the spacing between vectors of $450.6\ \mathrm{\mu m/vector}$. As shown in Fig.~\ref{fig2} a--b, highlighted with red dotted line, the region near the closure cannot yield reliable vectors due to the unsteady closure. Such spurious vectors (closure region of Fig.~\ref{fig2}b) are generated due to the non-uniform reflection of the light sheet that illuminates the unsteady closure. 
To cope with the cavity length fluctuations (Fig.~\ref{fig2}d) and the spurious vectors beyond the closure, we have set a threshold based on the standard deviation of velocity vectors (Fig.~\ref{fig2}c) which results in the reliable measurement field of view of PIV covering 85\% of the cavity length. 
To acquire the mean internal flow field of OC from the instantaneous flow fields with fluctuating length, the horizontal axis of each instantaneous flow field is normalized by the corresponding cavity length and then the normalized flow fields are averaged. 

\begin{figure*}[ht]
  \centering
  
  \includegraphics[width=1\textwidth]{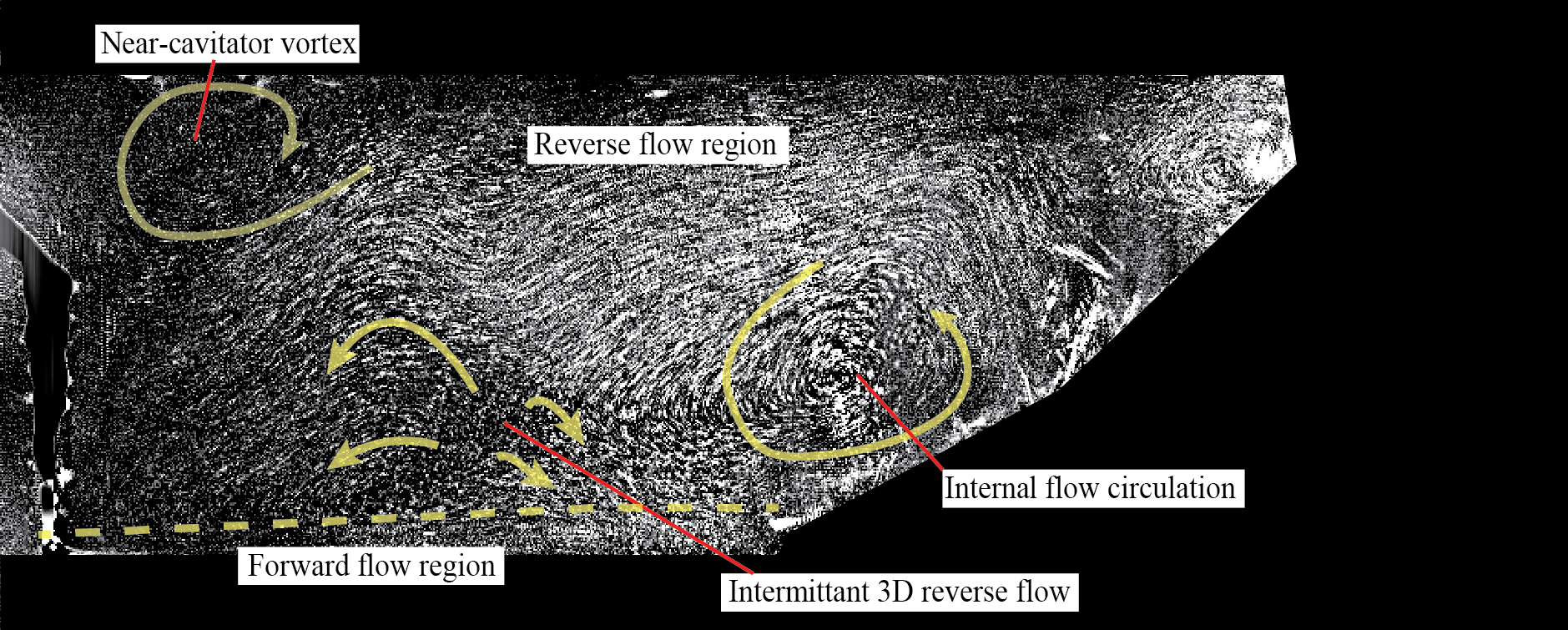}
  
  \caption{Tracer streak image of case 1 ($Fr=3.0$ and $C_Q=3.20\times 10^{-3}$) showing the internal flow of OC. Clockwise near-cavitator vortex, counter-clockwise internal flow circulation vortex, and intermittent transverse flow are commonly observed in tracer streak images.}
\label{fig4}
\end{figure*}

Due to the extended cavity length of TBC, the internal flow of TBC is recorded with separate FOVs of 1024$\times$1024 pixels, with the spatial interval of $3 \ \mathrm{cm}$ to ensure the presence of the overlapping region between FOVs. 
Figure~\ref{fig3}a shows the sample image of the tracer field illustrating the measurement field of view and image quality. The corresponding instantaneous velocity field is presented in Fig.~\ref{fig3}b. 
For clarity, only one in two vectors is plotted in both $x$ and $y$ directions.
For TBC, the interrogation window for PIV starts from 96$\times$96 pixels with a 50\% overlap to 32$\times$32 pixels with 50\% overlap. The resulting parameters yield the spacing between vectors of $524.6\ \mathrm{\mu m/vector}$. To obtain the entire mean flow of TBC from separate velocity vector fields (Fig.~\ref{fig3}b), each FOV is combined with a weighted average of distance for the overlapping regions. The heights of adjacent FOVs are matched to yield the minimum mean squared error between the two vector matrices in each overlapping region.

For all experimental conditions, temporal resolutions are designed to ensure that the smallest interrogation window (32$\times$32 pixels) captures forward moving tracers driven by an air-water interface within 10 pixel displacement per frame. The sampling time for each experimental condition is chosen to be long enough to ensure statistical convergence. Here we assume that the recorded data are statistically converged when the sampling time is more than three times greater than the time required for the reverse flow to be fully advected from the closure to the cavitator. The reverse flow velocity for calculating the sampling time was empirically determined by preliminary experiments, being roughly a half of the forward flow velocity driven by the air-water interface.

\section{Results}\label{sec:Results}

\subsection{OC internal flow field}\label{sec:OC}
Flow visualization based on tracer streak images provide a qualitative view on the flow features of the OC internal flow. The tracer streak images are generated with a maximum intensity projection of stack of 10 tracer images. 
As shown in Fig.~\ref{fig4} (flow visualization of case 1) and supplemental video S1 (flow visualization of case 2), the OC internal flow field exhibits distinct regions: the forward flow region, reverse flow region, near-cavitator vortex, and internal flow circulation vortex near the closure. 
The forward flow region is driven by the moving air-water interface at the speed of a free stream velocity. The reverse flow region, driven by an adverse pressure gradient, covers a large portion of the internal flow field, and the yellow dashed line marks the boundary between the forward and reverse flow regions. A gradual increase of the internal boundary layer thickness is observed along the air-water interface. 
Near-cavitator vortex refers to the region of the flow circulation near the cavitator, which is a result of the interaction between forward-moving ventilation jet and the reverse flow. 
It is noteworthy that the shape and location of the near-cavitator vortex resembles the \emph{secondary vortex} in a single phase flow past a backward-facing step, but its property relies on the ventilation coefficient ($C_Q$), which depends both on $Q$ and $U$. 
The internal flow circulation vortex near the closure is formed by the interaction between the forward and reverse flow. 
While \citet{qin2019IJMF} hypothesized that the gas leakage mechanism of OC (i.e., the dispersal of small bubbles from the air pocket) is primarily driven by the external recirculation vortex, our internal flow visualization reveals that the dynamics of internal flow circulation also contributes to the gas leakage of OC. 
Specifically, the presence and the movement of the internal flow circulation vortex influence the stability of the air-water interface, which ultimately leads to the intermittent shed-off of the air pocket at the closure. 
Remarkably, as shown by the short tracer streak patterns in the Fig.~\ref{fig4} and supplemental video S1, our internal flow visualization also suggests the presence of intermittent transverse flow in the reverse flow region. Such flow may be caused by the non-uniform pressure distribution along the spanwise direction associated with the unsteady cavity closure.

\begin{figure*}[t]
  \centering
  
  \includegraphics[width=1\textwidth]{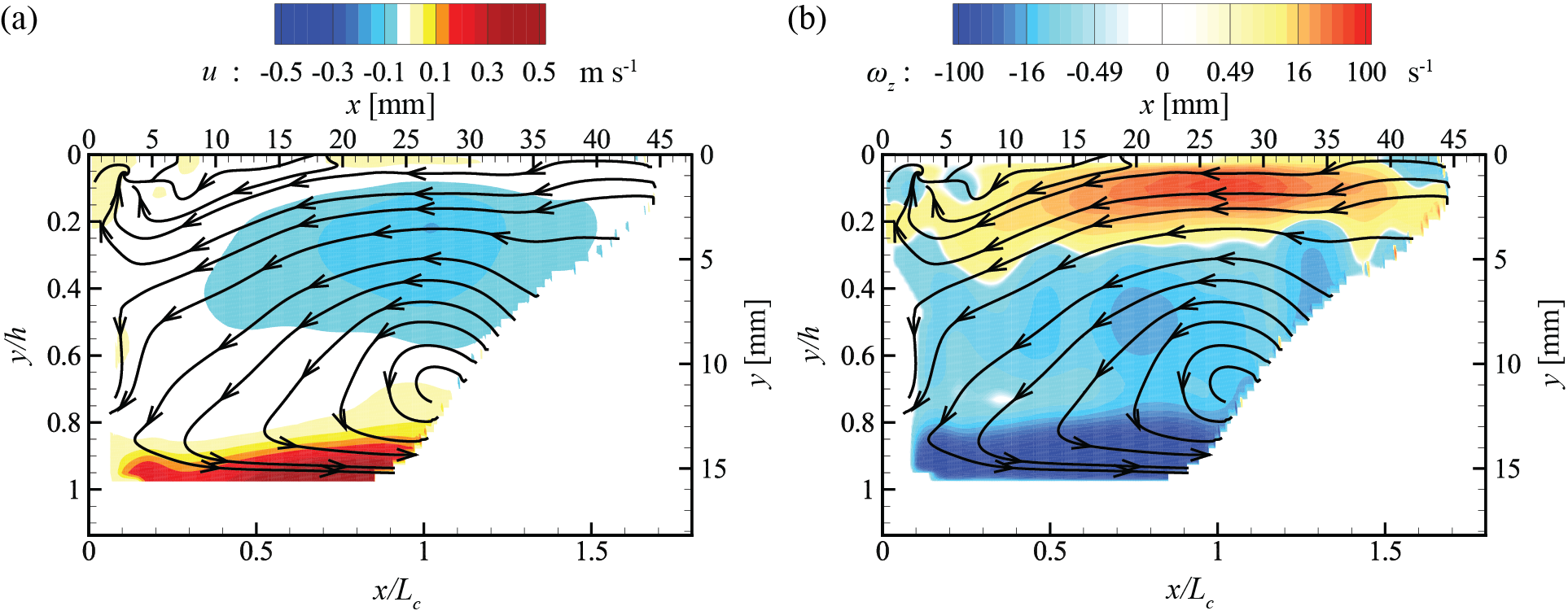}
  
  \caption{(a) Streamwise velocity and (b) vorticity contour maps superimposed with streamlines for case 1 ($Fr=3.0$ and $C_Q=3.20\times 10^{-3}$).}
\label{fig5}
\end{figure*}

\begin{figure}[t]
  \centering
  
  \includegraphics[width=0.48\textwidth]{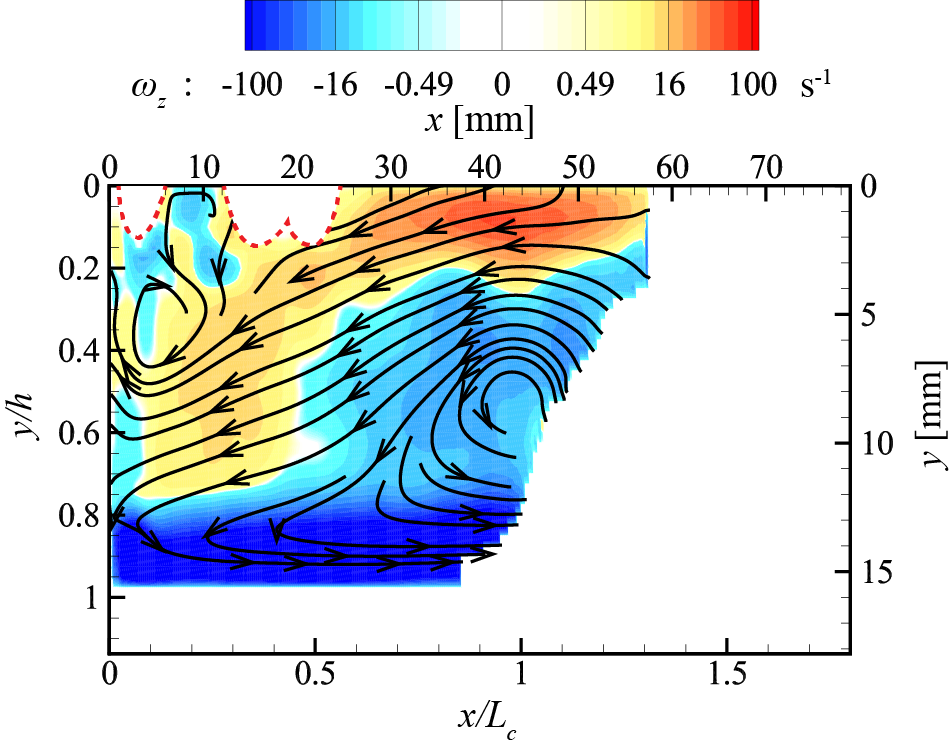}
  
  \caption{Vorticity contour maps superimposed with streamlines for case 2 ($Fr=3.0$ and $C_Q=3.20\times 10^{-3}$). The red dotted lines indicate the region that cannot be resolved due to the condensed droplets that block the optical access as mentioned in Section~\ref{sec:Methods}.3.}
\label{fig6}
\end{figure}

To provide a quantitative view of the internal flow field, the mean internal flow fields under constant $U$ with different $Q$ are investigated as follows. 
The mean flow fields are generated by following the procedure mentioned in Section~\ref{sec:Methods}.3. 
Figure~\ref{fig5} demonstrates the mean internal flow field for case 1 with the streamwise velocity and vorticity contours superimposed with streamlines.
The streamline pattern of case 1 clearly differentiates the distinct flow regions: the forward and reverse flow regions, near-cavitator vortex, and the internal flow circulation vortex. 
The distance from the cavitator to the center of the internal flow circulation vortex coincides with the mean cavity length ($x=L_c$) where the air film (air-water interface) is dispersed into bubbles, suggesting the role of internal flow circulation vortex in the gas leakage mechanism of OC. 
The increase of internal boundary layer thickness with streamwise location can be observed from the streamwise velocity contour (Fig.~\ref{fig5}a), which is consistent until the streamlines become disconnected at the closure. 
The vorticity contour map (Fig.~\ref{fig5}b) shows two strong shears in the mean flow field, one corresponding to the forward flow that is dragged by the moving air-water interface and another one corresponding to the reverse flow with high-vorticity shear near the ceiling. 
In addition, the vorticity map clearly shows the presence of the near-cavitator vortex and the internal flow circulation vortex, with the latter being significantly stronger and having an opposite sigh of vorticity compared to the former. This trend is consistent with the general pattern observed from the instantaneous sample (Fig.~\ref{fig4}). However, it is worth noting that the near-cavitator vortex on the mean vorticity contour map appears to be a little distorted and has a weaker strength compared to the observation from the instantaneous sample. Such discrepancy between the instantaneous and the averaged results suggests a fluctuation of the centroid location and the strength of the near-cavitator vortex potentially associated with the unsteadiness of the internal flow.

Figure~\ref{fig6} examines how the change of $C_Q$ due to an increased $Q$ affects the internal flow of OC.
It is important to note that during the OC experiments a small portion of the ceiling is occasionally covered by water droplets generated due to the unavoidable breakup of the air-water interfaces. Although the presence of these droplets blocks the tracer signal in some small regions near the ceiling, we have ensured that their presence does not distort the laser sheet within the recorded time sequence. Such regions are excluded during the PIV calculation to eliminate the spurious vectors associated with the presence of the droplets. Such regions are marked with red dotted lines in Fig.~\ref{fig6}.
As the figure shows, besides an increase of the mean cavity length as reported by \citet{qin2019IJMF}, the general features of internal flow of OC, i.e., the presence of forward and reverse flows as well as the near-cavitator and internal flow circulation vortices, remain unchanged with increasing $Q$. 
Particularly, the center of the internal flow circulation vortex stays around $x=L_c$ regardless of ventilation rates. However, the near-cavitator vortex shows an increase in strength and size with increasing $Q$. Additionally, the region of internal flow circulation appears to be more confined to the region near the closure.

\begin{figure*}
  \centering
  \includegraphics[width=1\textwidth]{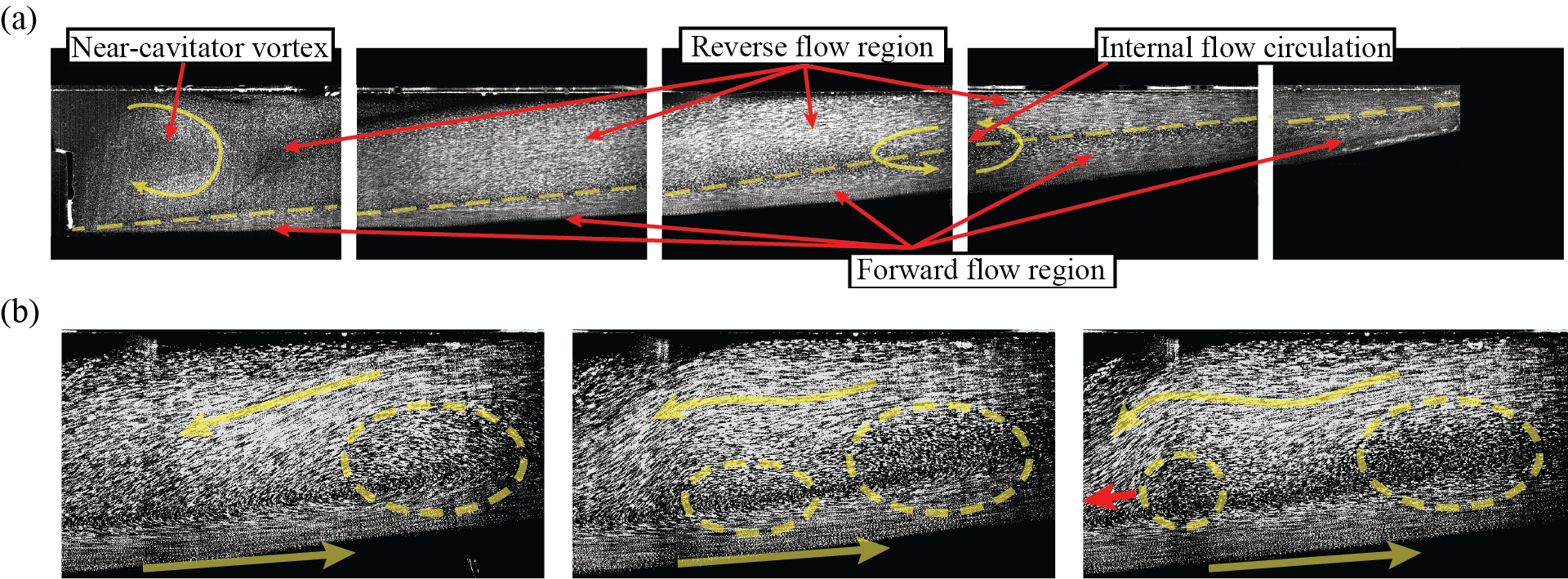}
  
  \caption{(a) Compilation of the tracer streak images at different streamwise locations case 3 ($Fr=\mathrm{1.6}$ and $C_Q=8.44\times 10^{-3}$) illustrating the presence of distinct flow regions inside TBC. Yellow dotted lines mark the boundary between forward and reverse flows. (b) Sequence of tracer streak images showing the presence of KH instability in the internal flow region. Yellow dotted circles mark the stagnant region of internal flows that is advecting upstream, providing an evidence of KH instability between forward and reverse flow regions. Such stagnant regions moving upstream causes the reverse flow to have an ocsillatory motion.
  }
\label{fig7}
\end{figure*}
\begin{figure*}
  \centering
  
  \includegraphics[width=1\textwidth]{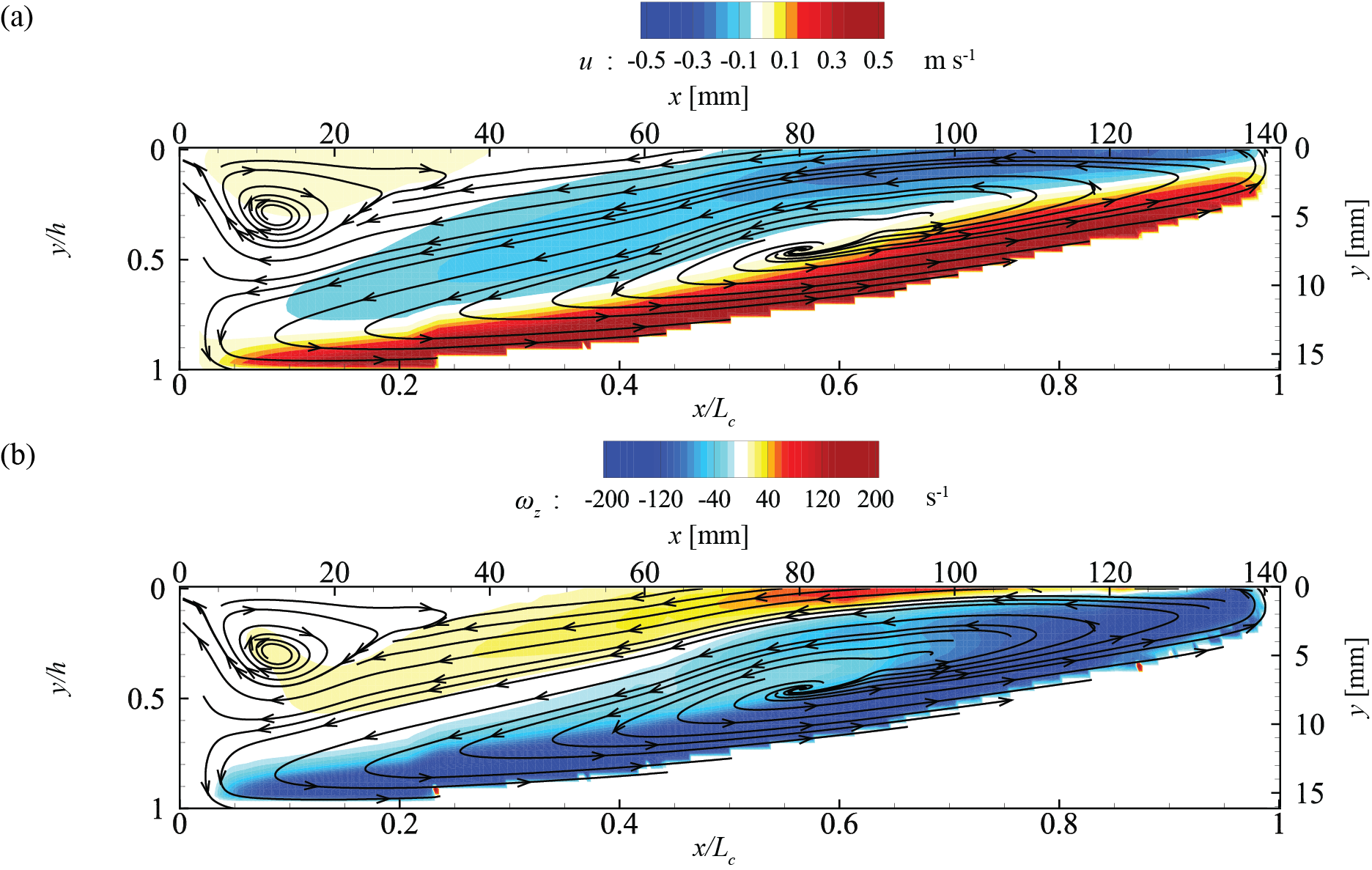}
  
  \caption{(a) Streamwise velocity and (b) vorticity contour map superimposed with streamlines of TBC case 3 ($Fr=\mathrm{1.6}$ and $C_Q=8.44\times 10^{-3}$).}
\label{fig8}
\end{figure*}

\begin{figure*}
  \centering
  
  \includegraphics[width=1\textwidth]{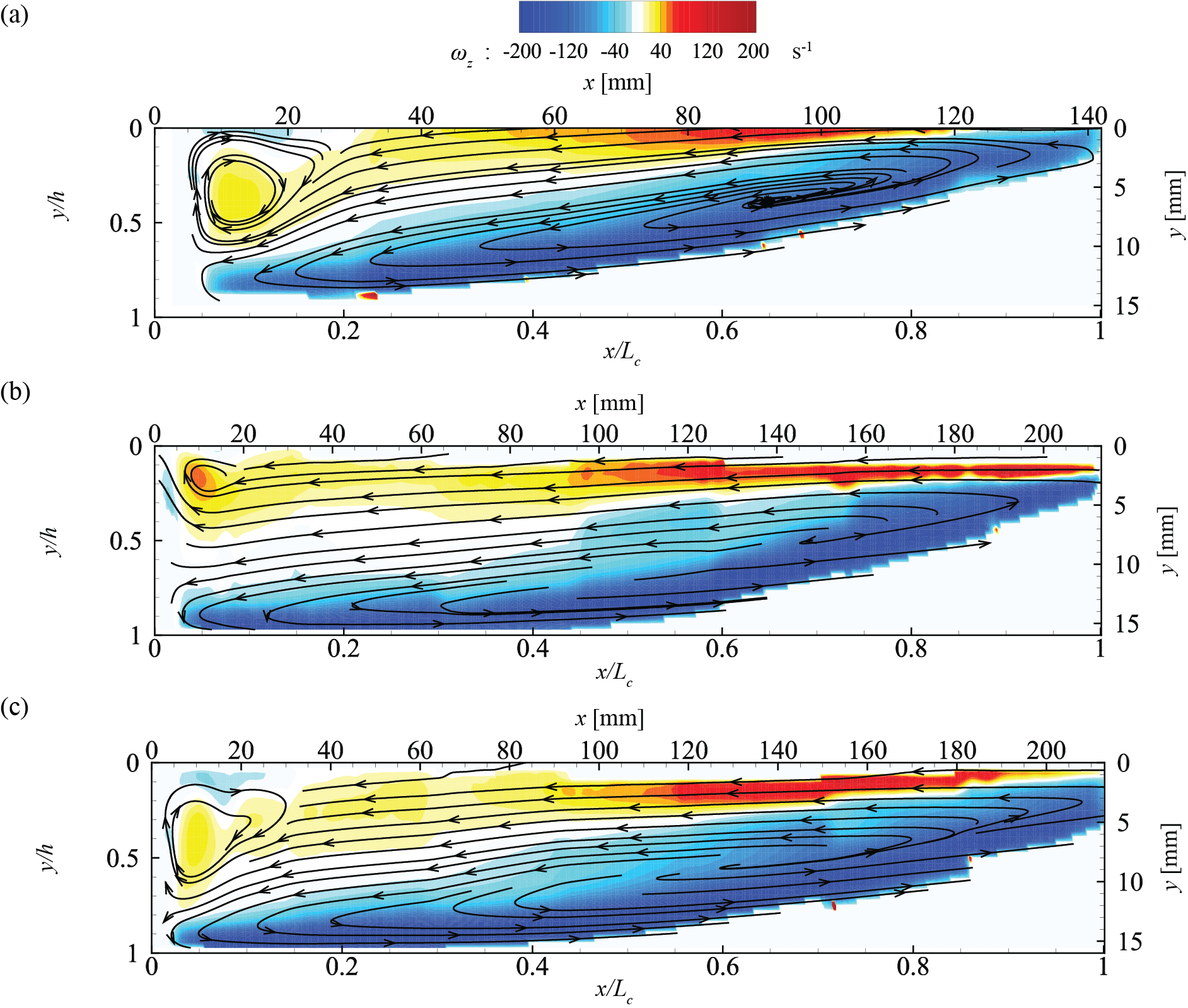}
  
  \caption{Vorticity contours of (a) case 4 ($Fr=1.6$ and $C_Q=4.22\times 10^{-2}$) that represents an increase in $Q$ from case 3, (b) case 5 ($Fr=2.0$ and $C_Q=6.85\times 10^{-3}$) that represents an increase in $Fr$ from case 3, and (c) case 6 ($Fr=2.0$ and $C_Q=3.43\times 10^{-2}$) that represents an increase in both $Fr$ and $Q$ from case 3. Each vorticity contour is superimposed with streamlines of the corresponding case.}
\label{fig9}
\end{figure*}

\subsection{TBC internal flow field}\label{sec:TBC}
Flow visualization with tracer streak images of TBC provides qualitative view on the internal flow structure at the center plane. Tracer streak images of TBC are generated following the same method as for the OC cases. Figure~\ref{fig7}a illustrates the snapshot of tracer streaks of each FOV that is characterized by case 3. 
Similar to OC internal flow, tracer streak images of TBC enable a clear observation of the forward and reverse flow regions, the near-cavitator vortex, and the internal flow circulation vortex. 
However, as shown from the supplemental videos S2--S6, the internal flow structure of TBC is more laminar-like and steady over time, compared to the instantaneous flow visualization of OC.
Particularly, the intermittent transverse flow occurring in the reverse flow region is no longer observed in TBC due to the absence of unsteady closure (see supplemental video S6).

However, as shown in Fig.~\ref{fig7}b, the reverse flow driven by the adverse pressure gradient is still unsteady, in spite of the steady cavity length with the air-water smoothly attached to the ceiling at the center $xy$-plane. The figure shows the stagnant flow regions marked by the short tracer streaks slowly advecting upstream and inducing wavy pattern of the reverse flow. Such unsteadiness suggests the presence of an instability in the internal flow region that affects the instantaneous flow structure of TBC. In particular, we believe that such instability is related to the Kelvin-Helmholtz (KH) instability, triggered by the strong velocity gradient across the interface of forward and reverse flows (see supplemental video S7 for clear visualization of such behavior).

\begin{table*}[ht]
\caption{List of experimental research on single phase flow separation over a backward-facing step with relevant expansion ratio ($ER$) and Reynolds number based on step height ($Re_h$).}
\label{table3}
    \begin{tabular}{ l l l l l }
    \hline\noalign{\smallskip}
      \thead{Author} & \thead{Technique} & \thead{$ER$} & \thead{$Re_h$} & \thead{$X_r/h$}\\
      \hline\noalign{\smallskip}
      \citet{beaudoin2004EJM}   & PIV \& dye injection  & $1.11$            & $50\--200$& $3\--9$\\
      \citet{goldstein1970JBE}  & Smoke injection       & $1.02\--1.07$     & $100\--520$& $0.021\times Re_h+2.13$ \\
      \citet{boiko2011EXIF}     & PIV                   & $1.004$           & $1060$    & $20$\\
      \citet{etheridge1978JFM}  & Anemometer            & $1.07$            & $3000$    & $5$\\
      \citet{kostas2002EXIF}    & PIV                   & $1.02$            & $4600$    & $4.8$\\
      \citet{scarano1999POF}    & PIV                   & $1.2$             & $5000$    & $6$\\
      \citet{nadge2014EXIF}     & PIV                   & $1.10$            & $20000$   & $6$\\
      \citet{ma2017POF}         & PIV                   & $1.05$            & $20000$   & $7.1$\\
      \citet{jovic1995EXIF}     & Laser-oil interferometer& $1.09$          & $6800$    & $5.35$\\
                                &                       & $1.2$  & $25500$ & $6.9$\\
\hline\noalign{\smallskip}
    \end{tabular}
    
\end{table*}

Mean flow fields of TBC are investigated to provide a quantitative view of the internal flows. 
The internal flow field of TBC is acquired by combining the mean flow fields of separate FOVs into a single flow field. 
Similar to the qualitative observation of TBC internal flow, the velocity contour map of case 3 (Fig.~\ref{fig8}a) shows the four distinct flow regions of TBC: the near-cavitator vortex, forward and reverse flow regions, and the internal flow circulation vortex.
Moreover, besides the substantial increase in cavity length compared to OC, a small forward flow region near the cavitator can be clearly observed from the velocity contour map, due to the relatively higher $C_Q$ for TBC cases.
The streamline pattern of TBC is different from that of the OC near the closure as the internal flow circulation vortex is fully connected, implying no interaction between internal and external flows at the center $xy$-plane that influences the stability of the air-water interface near the closure. 
The internal boundary layer thickness of TBC grows with streamwise direction to a certain extent, but then decreases due to the thinning of the cavity thickness near the closure. 
In addition, near the closure, reverse flow velocity appears to have the highest magnitude, and the slight attenuation of the forward flow is also observed. Such behaviors suggest the (spanwise) bifurcation of the internal flows due to the reduction of the cross-sectional area.
The vorticity contour map from Fig.~\ref{fig8}b shows the two strong shear regions with high vorticity near the air-water interface and near the ceiling, respectively. 
In addition, unlike for OC, the near-cavitator vortex is more clear from the vorticity contour map. An observation of the stronger vorticity at the near-cavitator vortex also indicates that the internal flow of TBC is more laminar-like compared to OC cases with appreciable ventilation jet near the cavitator.
The strength of the near-cavitator vortex is still weaker compared to that of the internal flow circulation vortex with opposite sign of vorticity.

Further investigation of the TBC internal flow is conducted with different $U$ and $Q$ to examine the effect of $C_Q$ on the internal flow structures of TBC. 
Figure~\ref{fig9}a presents the vorticity contour map superimposed with streamlines of case 4, which represents the increased $Q$ (from $1\ \mathrm{SLPM}$ to $5\ \mathrm{SLPM}$). As shown in the figure, the cavity length of TBC is invariant under the change of the ventilation rate, which is consistent with the findings reported by \citet{qin2019IJMF}. However, the internal flow structure near the cavitator differs with case 3 and the amplification of the near-cavitator vortex with an increase in $Q$ is clearly observed. The internal flow circulation vortex also exhibits strengthened vorticity with an increase in $Q$.
Furthermore, the effect of $C_Q$ on the TBC internal flow with higher $U$ (case 5) is visualized by the shrinkage in the size of the near-cavitator vortex (Fig.~\ref{fig9}b). Figure~\ref{fig9}c shows the consistency of the effect of $C_Q$ on the internal flow of TBC under different $U$ and $Q$. The size of the near-cavitator vortex of case 6 (Fig.~\ref{fig9}c) is enlarged compared to case 5 (higher $Q$ with constant $U$) similar to the aforementioned comparison between different $Q$ (i.e., Fig.~\ref{fig8}b and Fig.~\ref{fig9}a), and decreased compared to case 4 (higher $U$ with constant $Q$) similar to the aforementioned comparison between different $U$ (i.e., Fig.~\ref{fig8}b and Fig.~\ref{fig9}b). The strength of internal flow circulation vortex is also observed to be strengthened with increasing $Q$. 
Disconnected isocontours in the vorticity field with increased $U$ suggests an increase in the unsteadiness of the internal flow structure that was previously discussed in Fig.~\ref{fig7}b.
The location of the internal flow circulation vortex, however, appears to be located in similar regions despite the change in $U$ or $Q$.

\begin{figure}[t]
  \centering
  
  \includegraphics[width=0.48\textwidth]{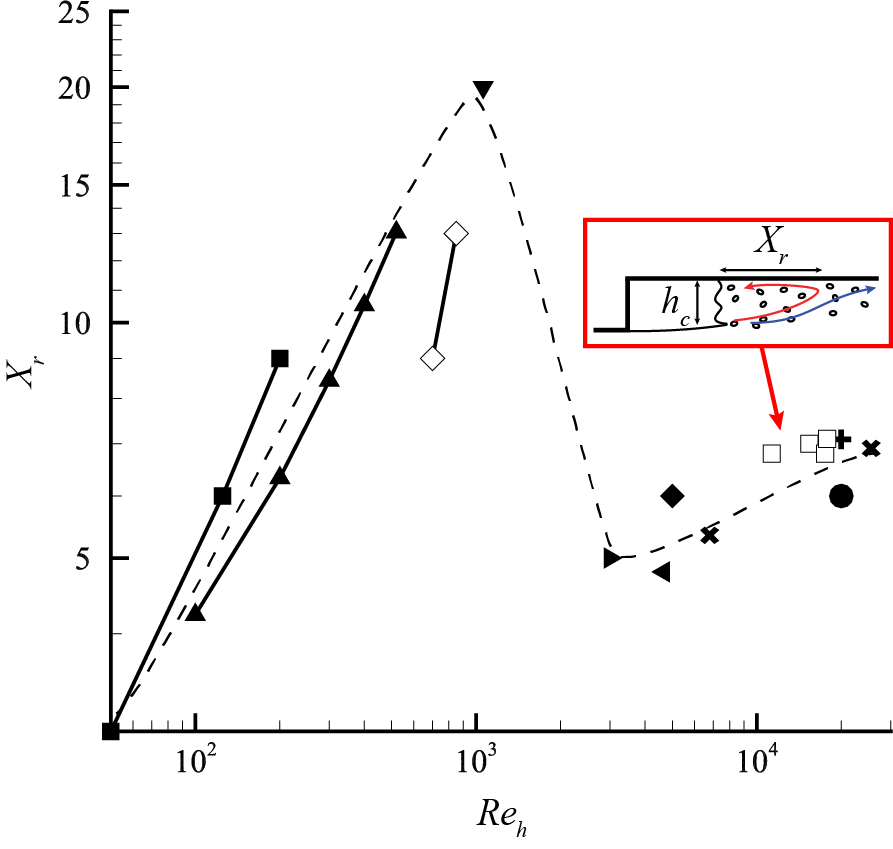}
  
  \caption{Reattachment lengths ($X_r$) from the literature of low expansion ratio and varying Reynolds number based on step height. Dotted line indicates the trend of the $X_r$ of the single phase flow separation with increase in $Re_h$. Symbols: $\blacksquare$, \cite{beaudoin2004EJM}; $\blacktriangle$, \citet{goldstein1970JBE}; $\blacktriangledown$, \citet{boiko2011EXIF}; $\blacktriangleright$, \citet{etheridge1978JFM}; $\blacktriangleleft$, \citet{kostas2002EXIF}; $\blacklozenge$, \citet{scarano1999POF}; $\bullet$, \citet{nadge2014EXIF}; $+$, \citet{ma2017POF}; $\times$, \citet{jovic1995EXIF}; $\square$, $X_r$ of external recirculation region of OC from \citet{qin2019IJMF}; $\diamond$, TBC length from the present study. Note that the $X_r$ of single phase flows are marked with filled symbols, and the data points corresponding to VPC studies of ours and \citet{qin2019IJMF} are marked with hollow symbols.}
\label{fig10}
\end{figure}

\subsection{Comparison with flow over a backward-facing step}\label{sec:Analogy}
To further understand the internal flow field of VPC, we provide a comparison of VPC internal flows with a single phase flow over a backward-facing step ($BFS$), which is referred to as $BFS$ flow hereafter.
Since $BFS$ flow is influenced by the expansion ratio ($ER$) according to the literature, only the studies with $ER<1.2$ are selected and are tabulated in Table~\ref{table3} for comparing with our VPC internal flow cases which uses a cavitator corresponding to a $ER$ of $1.09$. Figure~\ref{fig10} illustrates the trend of the reattachment length ($X_r$) of $BFS$ flow upon change in $Re_h$. As shown in the figure, for $BFS$ flow, its $X_r$ increases approximately linearly when the flow is laminar, from $3h$ to $20h$. Then it drops abruptly to about $5h$ as the flow transitions from laminar to turbulent at a critical $Re_h$ of around 2000. With further increase in $Re_h$, $X_r$ grows at a slower rate in comparison to that in the laminar regime, with a value ranging from $5h$ to $7h$ under $Re_h$ of $3000$ to $25500$.

To compare with $BFS$ flow, a proper definition of $Re_h$ for VPC internal flow needs to consider the properties of both air and water phases since it is the interplay between the two phases that dictates the overall behavior of the cavity.
Remarkably, as the schematic shown in Fig.~\ref{fig10}, in OC regime, the $X_r$ of the external recirculation region outside the cavity and its variation upon $Re_h$ is comparable to that of $BFS$ flow in turbulent regime under the same $Re_h$ when it is determined using the water viscosity and the cavity thickness at the closure of OC ($h_c$). This result suggests that the external recirculation region of OC is turbulent, consistent with the assessment from \citet{qin2019IJMF}. 
Moreover, such turbulent external recirculation results in a fluctuation of the cavity length of OC (evidenced in Fig.~\ref{fig2}d) up to one step height ($h$). 
This phenomenon resembles the fluctuation of the $X_r$ in the turbulent $BFS$ flow within a range of $3h$ \citep{schafer2009JFM,ma2017POF}, suggesting a strong coupling of the internal air flow with the water flow in the external recirculation region. 
For turbulent $BFS$ flow, the fluctuation of the $X_r$ is attributed to the onset of KH instability that triggers the flapping motion in the separated shear layer \citep{ma2017POF}. 
Accordingly, the presence of KH instability in the separated shear layer extended from the cavity interface may be responsible for the cavity length fluctuation of OC. 
In contrast, in TBC regime, the cavity length and its growth trend upon $Re_h$ is found to match those of the $X_r$ of laminar $BFS$ flow under the same $Re_h$ based on the viscosity of air. 
Such matching implies that the cavity behavior of TBC is dictated by the internal air flow and tends to be decoupled from the influence of external flow disturbance as the air-water interface is smoothly attached to the ceiling at the closure with an absence of the external recirculation region.

In accordance with the cavity geometry, the internal flow fields of VPC are also compared with the $BFS$ flows. 
Specifically, for OC, its internal mean flow resembles the turbulent BFS flow in terms of flow structures. For instance, the behavior of the internal flow circulation vortex and near-cavitator vortex of OC are similar to the recirculation vortex and secondary vortex of $BFS$ flow, respectively. 
In addition, both flow fields show a maximum reverse flow velocity occurring at the center of the circulation vortex, i.e., the internal flow circulation vortex of OC (Fig.~\ref{fig5}a) and the recirculation vortex of turbulent $BFS$ flow \citep{ma2017POF}. 
To provide a better comparison between the flow structures, we introduce the velocity ratio of the forward flow and the reverse flow along the same velocity profile at each streamwise location. It should be noted that the maximum of this ratio ($r_{FR,max}$) yields $0.15$ for OC internal flows and $0.14$ for turbulent $BFS$ flow \citep{ma2017POF}.
Moreover, the unsteady behaviors of OC internal flow resemble those of turbulent $BFS$ flow. 
As it can be observed from supplemental video S1, the internal flow circulation vortex of OC exhibits intermittent breakdown and its advection towards upstream. 
Such behavior is referred to as the convective instability, since the amplified disturbance diminishes in its origin and convects to other locations \citep{Huerre1990ARFM}. 
Remarkably, such convective instability also occurs in the turbulent $BFS$ flow but is manifested as KH instability in the separated shear layer shown as intermittent vortical structures advecting downstream.
As a result, it is conceivable that the unsteady behavior of the internal flow circulation vortex may be influenced by the presence of the unsteadiness within the separated shear layer in the external recirculation region of OC, which provides further support to the strong coupling of the internal and external flows that determines the cavity length fluctuation of OC.

Correspondingly, the internal flow of TBC shows some similarities to the $BFS$ flow in the laminar regime. 
As observed in the supplemental videos S2--S6, the internal flow structure of TBC is laminar-like, despite some unsteadiness is observed within the reverse flow region (see Fig.~\ref{fig7}b). 
Such unsteadiness may be attributed to the KH instability triggered by the strong velocity gradient between forward and reverse flows, as mentioned in Section~\ref{sec:TBC}, which has not been reported in the laminar $BFS$ flow studies \citep{goldstein1970JBE,boiko2011EXIF}. 
Such velocity gradient can be characterized by $r_{FR,max}$ defined earlier, which turns out to be roughly $0.40$ for TBC cases, considerably higher than $r_{FR,max}$ of the laminar $BFS$ flow that has been reported as around $0.05$ \citep{boiko2011EXIF}.
Such high $r_{FR,max}$ (see Fig.~\ref{fig8}a) near the closure, mainly due to the reduced cross-sectional area and the bifurcation of the internal gas flow, results in stronger velocity gradient for TBC that triggers the KH instability.
It is noteworthy that the location of the maximum reverse flow velocity is not directly comparable to the flow structure of the laminar $BFS$ flow. However, such phenomenon clearly points out the difference from that of OC, where its maximum reverse flow velocity and its location, and their maximum velocity ratios are comparable to those of the turbulent $BFS$ flow.

Finally, the response of the VPC to changing free stream velocity and ventilation rates can also be better understood through a comparision with the behavior of BFS flow upon changing in $Re_h$. With increasing $U$ under fixed ventilation, the cavity length of TBC first increases and then drop abruptly as the cavity transitions from TBC to OC. 
This phenomenon is similar to that occurs in the change of $X_r$ of $BFS$ flow as the flow field transitions from laminar to turbulent upon increasing $Re_h$. 
This comparison implies the change of cavity length in VPC is strongly associated with the transition of internal gas flow from laminar-like to highly unsteady, and the delay of such transition may help sustain the presence of TBC upon changing external flows. 
Our results suggest that such delay, or the control of state of the internal flow, can be achieved through ventilation. 
Particularly, it has been found that the internal pressure of VPC measured at the cavitator grows with ventilation \citep{qin2019IJMF}, indicating a growth of favorable pressure gradient with increasing $Q$. Such behavior explains the transition from OC to TBC with increasing $Q$ under fixed $U$ due to a laminarization process of turbulent flow with favorable pressure gradient \citep{patel1968JFM}. 
Further increase in $Q$ beyond the transition can help suppress the growth of instability within the shear region that affects the internal flow structure, allowing the cavity to be sustained in its TBC regime even at the higher $Re_h$ where the laminar $BFS$ flow may transition into turbulent.

\section{Conclusions}\label{sec:Conclusions}
In this study, we conducted flow visualization and particle image velocimetry experiments using fog particles as tracers to provide the internal flow characterization of a ventilated partial cavitation. Various experiments are conducted at two different cavity regimes, i.e., open cavity (OC) and two-branch cavity (TBC) under various free stream velocity ($U$) and ventilation rates ($Q$). 
Instantaneous flow visualization and mean flow field show the presence of general internal flow structures for both OC and TBC, i.e., forward flow region near the moving air-water interface, reverse flow region occupying the large portion of the internal flow, and vortical structures (near-cavitator vortex and internal flow circulation vortex) swirling in opposite directions. 
The comparison between instantaneous and mean flow fields show that the OC internal flow is highly unsteady, and its internal flow circulation vortex breaks down intermittently and advects upstream. 
In contrast, the TBC internal flow is laminar-like despite some unsteadiness within the reverse flow region associated with the Kelvin-Helmholtz (KH) instability triggered by the strong velocity gradient at the interface between the forward and reverse flows.
The comparison between VPC internal flows and the single phase flow past a backward-facing step ($BFS$ flows) further reveals the cause of the aforementioned unsteadiness in the internal flow regions. For OC, it is suggested that the internal flow in air phase is highly coupled with the external recirculation region which is in turbulent water phase, and the unsteadiness of the internal flow structure may be influenced by the convective instability from the external recirculation region. 
For TBC, the internal flow structure resembles a laminar $BFS$ flow, despite the presence of the unsteadiness within reverse flow due to the KH instability. However, such unsteadiness in TBC internal flow does not contaminate the overall flow structure. 
Based on such comparison, explanations are made on the cavity behavior and the regime transitions under the change of $U$ and $Q$. 
As $U$ increases from TBC, internal flow structure gradually becomes unstable and exhibits an increased unsteadiness within the reverse flow region, similar to the onset of convective instability within the separated shear layer of $BFS$ flow that triggers the transition of its regime from laminar to turbulent. Further increase in $U$ results in highly unstable internal flow with the cavity regime transitioning from TBC to OC, and its internal flow exhibits strong coupling with the external flow which is in turbulent regime. 
As $Q$ increases from OC, in combination with favorable pressure gradient, highly unstable internal flow becomes more laminar which ultimately changes its regime to TBC.
Correspondingly, the cavity length of OC grows due to an increased role of air phase until the internal flow of OC becomes decoupled with the external flows.
Further increase in $Q$ beyond the transition suppresses the growth of perturbation from the KH instability in the strong shear region, and the cavity sustains in TBC regime even at higher $Re_h$ where the laminar $BFS$ flow may transition into turbulent.

Our measurements reveal the distinct flow structures of VPC under different cavity regimes and demonstrate the role of internal flow on the cavity geometry and regime transitions. 
In addition, our findings suggest that the ventilation control can potentially stabilize the cavity in the TBC regime by delaying its internal flow transition from laminar-like to highly unsteady. 
Therefore, to further elucidate the effect of ventilation on the dynamics of VPC, it is required to scope into the gas leakage mechanism associated with cavity transition and internal flow changes. However, such investigation requires 3D flow characterization for unsteady closure of OC and internal flow bifurcation of TBC, which cannot be performed with the current planar PIV measured only at the single spanwise location. 
For instance, our measurements characterize the internal flows using planar PIV at the center plane.
But to gain a better understanding of the unsteady closure of OC and the internal flow bifurcation of TBC, we recommend future works to measure and examine the three-dimensional flow characteristics.
In addition, our current study is limited in the relatively low range of $Fr$ due to the restricted length of the test section of the water tunnel. As a result, our study does not include VPC in higher $Re_h$ range for TBC where one can examine how the ventilation influence the unsteadiness of TBC internal flow.
This limitation can be overcome by using a smaller step height for the backward-facing step cavitator in the future experiments.



%
%


\begin{acknowledgements}
This work is supported by the Office of Naval Research (Program Manager, Dr T. Fu) under grant no. N000141612755.
\end{acknowledgements}

%
%

\bibliographystyle{spbasic}      


%
%

\end{document}